\documentclass[preprint,aps,amsmath,nofootinbib,amssymb,showkeys]{revtex4-1}
\usepackage{verbatim,graphics,graphicx,color,slashed,textcomp}
\usepackage{ulem} % this one is for cross out text
\usepackage[colorlinks=true,
            linkcolor=red,
            urlcolor=blue,
            citecolor=blue]{hyperref}

\graphicspath{{figures/}}

\newcommand{\eV}{\mathrm{eV}}

\newcommand{\GeV}{\mathrm{GeV}}
\newcommand{\TeV}{\mathrm{TeV}}

\newcommand{\eff}{\textrm{eff}}
\newcommand{\planck}{\texttt{Planck}}

\begin{document}

\title{Residual Non-Abelian Dark Matter and Dark Radiation}
\author{P. Ko}
\affiliation{Korea Institute for Advanced Study, Seoul 02455,  Korea}
\affiliation{Quantum Universe Center, KIAS, Seoul 02455,  Korea}
\author{Yong Tang}
\affiliation{Korea Institute for Advanced Study, Seoul 02455,  Korea}
\affiliation{Department of Physics, Faculty of Science, The University of Tokyo, Bunkyo-ku, Tokyo 133-0033, Japan}
\date{\today}

\begin{abstract}
We propose a novel particle physics model in which vector dark matter (VDM) and dark 
radiation (DR) originate from the same non-Abelian dark sector. We show an illustrating 
example where dark $SU(3)$ is spontaneously broken into $SU(2)$ subgroup by the nonzero 
vacuum expectation value (VEV) of a complex scalar in fundamental representation of $SU(3)$. 
The massless gauge bosons associated with the residual unbroken $SU(2)$ constitute 
DR and help to relieve the tension in Hubble constant measurements between {\tt Planck} 
and Hubble Space Telescope. In the meantime, massive dark gauge bosons associated with the broken generators 
are VDM candidates. Intrinsically, this non-Abelian VDM can interact with non-Abelian DR 
in the cosmic background, which results in a suppressed matter power spectrum and leads to 
a smaller $\sigma_8$ for structure formation. 
\end{abstract}
\maketitle

\section{Introduction}\label{sec:intro}
It has been well established that about $25\%$ of energy density in our Universe is made of 
non-baryonic dark matter (DM). From the perspective of particle physics, 
weakly-interacting massive particle (WIMP) is one of the nicely motivated candidates. 
In WIMP scenario, DM is in thermal equilibrium with standard model (SM) particles at high 
temperature and freezes out at later time. Such an optimistic framework has triggered 
enthusiastic DM searches in direct, indirect and collider detection experiments. 
However, we have to admit that so far all confirmed evidence for DM is only from gravitational interaction, which leaves  wide possibilities for DM's particle identities.  

Recently, there are renewed interests in interacting DM--DR models~\cite{Bringmann:2013vra, Boehm:2014vja, Ko:2014bka, Chu:2014lja, Buen-Abad:2015ova, Lesgourgues:2015wza, Tang:2016mot, Bringmann:2016ilk, Ko:2016uft} which could have distinguishing effects on large scale structure. Depending on the DM--DR interactions, these effects can be similar to baryonic acoustic oscillation or dramatically different. Motivations for such models are  at least twofold. 
One is that the DR component could help to resolve the conflict between \planck
~\cite{Planck:2015xua} and Hubble Space Telescope (HST) data~\cite{Riess:2016jrr}. 
The other is that interaction between DM and DR can give a smaller $\sigma_8$ for structure 
growth, suggested by low redshift measurements, such as weak lensing survey CFHTLenS~\cite{Heymans:2012gg}. These tensions have stimulated various investigations on cosmological models~\cite{Pourtsidou:2016ico, DiValentino:2016hlg, Qing-Guo:2016ykt, Archidiacono:2016kkh, Wyman:2013lza, Zhang:2014lfa, Lesgourgues:2015wza, Ko:2016uft, DiValentino:2016ucb}. 

In this paper, we propose a new scenario where DM and DR have the same origin from a 
single Yang-Mills dark sector, unlike early attempts where DM and DR have different identities~\cite{Bringmann:2013vra, Boehm:2014vja, Ko:2014bka, Chu:2014lja, Buen-Abad:2015ova, Lesgourgues:2015wza, Tang:2016mot, Bringmann:2016ilk, Ko:2016uft}. 
In our framework presented below, a non-Abelian gauge group is spontaneously 
broken into its non-Abelian subgroup. The massless gauge boson associated with 
the residual subgroup constitutes  non-Abelian DR, while other massive gauge 
bosons make  non-Abelian VDM candidates. Naturally, VDM can interact with DR 
through the original Yang-Mills gauge interactions, inducing some observable effects on cosmology and astrophysics.  

This paper is organized as follows. In Sec.~\ref{sec:su3}, we start with an explicit example 
where dark  $SU(3)$ is broken to its subgroup $SU(2)$ by nonzero VEV of a complex scalar belonging 
to the fundamental representation of $SU(3)$. Then we generalize to dark $SU(N)$ that is broken 
into $SU(N-1)$, and give a brief proof why the massive gauge bosons are stable and therefore 
make good DM candidates. Next in Sec.~\ref{sec:constraint}, we discuss some phenomenologies 
and constraints on such a class of models, especially on DM--DR scattering, DM self-interaction 
and DR's contributions to $N_\eff$. Then in Sec.~\ref{sec:relic}, we estimate how DM's relic density can be satisfied with freeze-in process. Later in Sec.~\ref{sec:numeric} we illustrate the effects on matter power spectra in the interacting DM--DR scenario. Finally, we give our conclusion.

\section{The Model}\label{sec:su3}
Let us begin with a simple, illustrating case with hidden $SU(3)$ broken into $SU(2)$. 
We consider a hidden sector complex scalar $\Phi$ that belongs to  the 
fundamental representation $SU(3)$ with the following Lagrangian:
\begin{equation}\label{eq:model1}
\mathcal{L}=-\frac{1}{4}F^a_{\mu\nu}F^{a\mu\nu} 
+ \left(D_\mu \Phi\right)^\dagger \left(D^\mu \Phi\right) - \lambda_\phi\left(|\Phi|^2-v^2_\phi/2\right)^2,
\end{equation}
where $F^a_{\mu\nu}=\partial_\mu A^a_\nu - \partial_\nu A^a_\mu +g f^{abc}A^b_\mu A^c_\nu$, covariant derivative $D_\mu$ is defined by $D_\mu \Phi=\left(\partial_\mu-igA_\mu^a t^a\right)\Phi$, and generators $t^a$s are normalized as $\textrm{Tr}[t^at^b]=\delta^{ab}/2$. For transparent presentation, we express the gauge field explicitly as
\begin{equation}
A_\mu \equiv A_\mu^at^a=\frac{1}{2}\left(
\begin{array}{ccc}
 A_\mu^3+\frac{1}{\sqrt{3}}A_\mu^8  & A_\mu^1-iA_\mu^2 & A_\mu^4-iA_\mu^5 \\
 A_\mu^1+iA_\mu^2 & -A_\mu^3+\frac{1}{\sqrt{3}}A_\mu^8 & A_\mu^6-iA_\mu^7 \\
 A_\mu^4+iA_\mu^5 & A_\mu^6+iA_\mu^7 & -\frac{2}{\sqrt{3}}A_\mu^8
\end{array}	
	\right).
\end{equation}
After $\Phi$ gets a non-zero vacuum expectation value (vev), in unitary gauge we would have
\begin{equation}
\langle \Phi \rangle =
\left( 0\;\; 0\;\;  \frac{v_\phi}{\sqrt{2}} \right)^T, \Phi=\left( 0\;\; 0\;\;  \frac{v_\phi+\phi\left(x\right)}{\sqrt{2}} \right)^T .
\end{equation}
Due to the spontaneous symmetry breaking by the above vacuum configuration, gauge bosons $A^{4,\cdots, 8}$ obtain masses from the interaction term $g^2 (A_\mu\Phi)^\dagger(A^\mu\Phi)$, 
\begin{equation}
m_{A^{4,5,6,7}}=\frac{1}{2}gv_\phi,\; m_{A^8}=\frac{1}{\sqrt{3}}gv_\phi,
\end{equation}
while gauge bosons $A^{1,2,3}$ associated with the unbroken gauge group $SU(2)$ are still massless.
 
\begin{figure}[t]
	\includegraphics[scale=0.7]{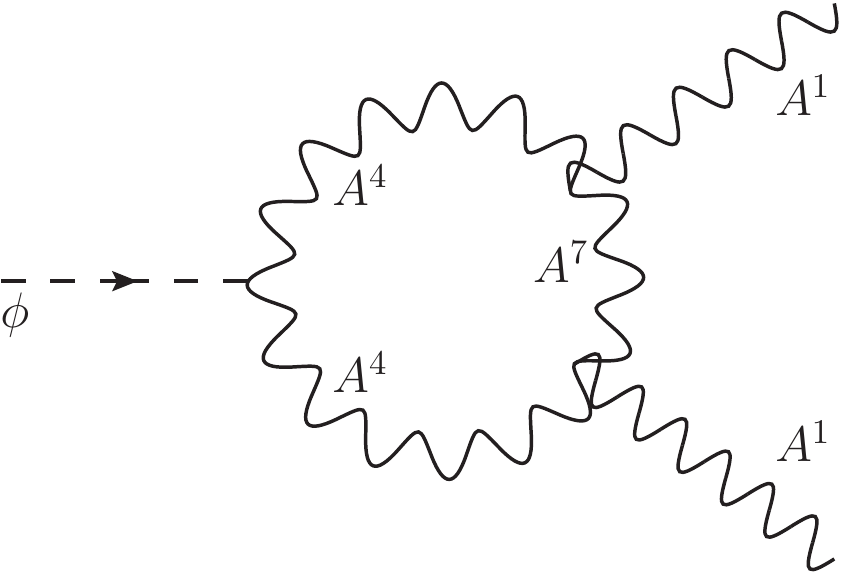} 
	\caption{An example Feynman diagram for decay $\phi\rightarrow A^a A^a$.\label{fig:decay}}
\end{figure}

One key feature we would like to point out is that the physical dark Higgs boson $\phi$ with mass $m_\phi=\sqrt{2\lambda_\phi}v_\phi$ couples to massive $A^{4,\cdots,8}_\mu$ as $\sim gm_A\phi A^m_\mu A^{m\mu}$ at tree level, but to massless $A^{1,2,3}$ much weakly as $\sim \dfrac{g^2}{16\pi^2 v_\phi}\phi F^{a\mu\nu}F^a_{\mu\nu}$ at one-loop level as Fig.~\ref{fig:decay}, which, as we shall show later, provides an alternative production mechanism for DM and DR rather than the usual thermal production.  

The interactions among $A^a_\mu$s are determined by $F^a_{\mu\nu}F^{a\mu\nu}$. 
For example, the vertex function for 
$A^a_\mu\left(k\right) A^b_\nu\left(p\right) A^c_\rho\left(q\right)$ is given by 
\begin{equation}\label{eq:3point}
	gf^{abc}\left[g^{\mu\nu}\left(k-p\right)^\rho+g^{\nu\rho}\left(p-q\right)^\mu+g^{\rho\mu}\left(q-k\right)^\nu\right],
\end{equation}
and the four-point $A^a_\mu  A^b_\nu  A^c_\rho A^d_\sigma$ by 
\begin{equation}\label{eq:4point}
-ig^2\left[f^{abe}f^{cde}\left(g^{\mu\rho}g^{\nu\rho}-g^{\mu\sigma}g^{\nu\rho}\right)+\left(b\leftrightarrow c,\nu\leftrightarrow \rho \right) + \left(b\leftrightarrow d, \nu\leftrightarrow \sigma \right)\right].
\end{equation}
The structure constants for $SU(3)$ are given by  
\begin{align}
f^{123}=1,f^{147}=-f^{156}=f^{246}=f^{257}=f^{345}=-f^{367}=\frac{1}{2}, f^{458}=f^{678}=\frac{\sqrt{3}}{2}.
\end{align}
All other $f^{abc}$s are zero if the indices $(abc)$ are not related the above ones by permutations. Now it is straightforward to check that $A^{4,5,6,7}$ always appears in pairs, equivalently having $Z_2$ symmetries. Because they have the same masses, $A^{4,5,6,7}$ would be stable and constitute as vector dark matter (VDM) candidates at renormalizable level. On the other hand, $A^8$ is lighter than 2$m_{A^4}$, so it can not decay into $A^{4,5,6,7}$ and is stable as well. In short, all massive $A^{4,\cdots ,8}$ are possible VDM  candidates
with two different masses, while massless $A^{1,2,3}$ are dark radiation (DR).

We can generalize the above discussions to the case in which dark $SU(N)(N>2)$ gauge 
symmetry is spontaneously broken into $SU(N-1)$ subgroup. We then have $2N-1$ massive 
gauge bosons as VDM candidates where their masses are given by 
\begin{equation}
m_{A^{\left(N-1\right)^2,...,N^2-2}}=\frac{1}{2}gv_\phi,\;m_{A^{N^2-1}}=
\frac{\sqrt{N-1}}{\sqrt{2N}}gv_\phi,
\end{equation}
and $N^2-2N$ massless gauge bosons as dark radiation. 
This can be proved by looking at the structure of $f^{abc}$. 
We divide the generators $t^{a}$ into two subsets, $a\subset[1,2,...,(N-1)^2-1]$ and 
$a\subset [(N-1)^2,...,N^2-1]$. Since $[t^a,t^b]=if^{abc}t^c$ for the first subset forms closed 
$SU(N-1)$ algebra, we have $f^{abc}=0$ if only one of the indices $(abc)$ is from the second 
subset. This also means that $f^{abc}\neq 0$ when either $a,b,c$ are all from the first subset, 
or at least two of them are from the second subset. If one index is $N^2-1$, then other two 
must be among the second subset to give no vanishing $f^{abc}$, because $t^{N^2-1}$ 
commutes with $t^{a}$ from $SU(N-1)$. Since $m_{A^{N^2-1}}<2m_{A^{\left(N-1\right)^2,...,N^2-2}}$, $A^{N^2-1}$ can not decay. 
Therefore, due to the interaction structure in Eqs.~(\ref{eq:3point}) and (\ref{eq:4point}), all $A^{\left(N-1\right)^2,...,N^2-1}$ are stable and can be VDM candidates. 
Generalization to more complicated breaking patterns or other gauge groups is similar. 
From now on, unless otherwise specified, we shall work in the $SU(3)\rightarrow SU(2)$ case and collectively denote $A^a$ with $a=1,2,3$ for $SU(2)$ and $A^m$ with $m=4,...,8$ for other massive ones. 

\begin{figure}[t]
	\includegraphics[width=0.8\textwidth, height=0.22\textwidth]{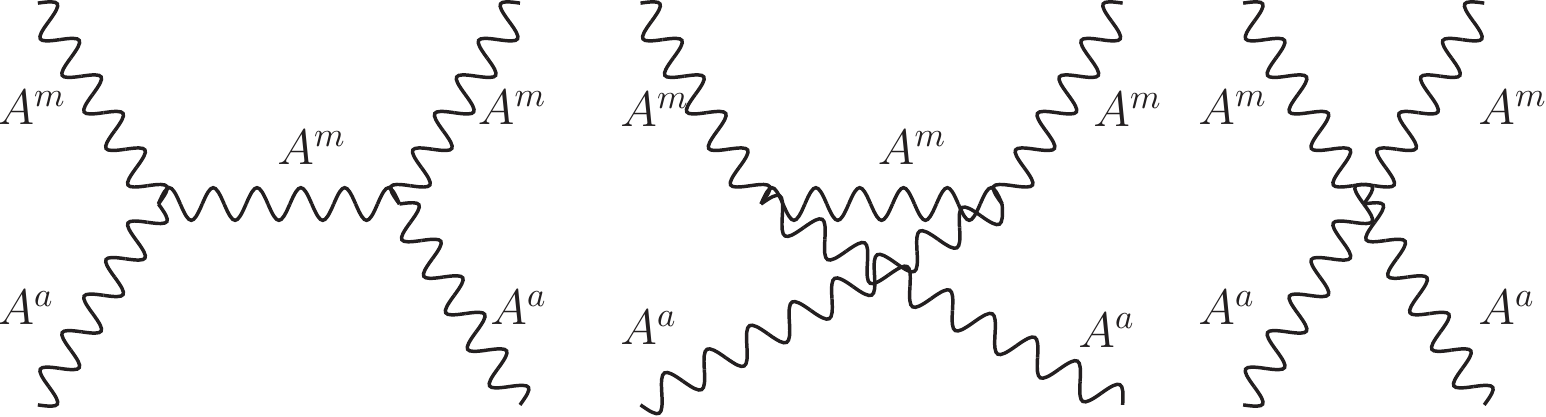} 
	\caption{Feynman diagram for DM-DR scattering in the $U(1)$ case.\label{fig:u1}}
\end{figure}

We should emphasize that the residual gauge group being non-abelian rather than $U(1)$ is crucial in this paper. 
Otherwise, the massless gauge boson is just ordinary DR~\cite{Baek:2013dwa, Khoze:2014woa, Kawasaki:2015lpf} but can not mediate sizable interaction between DM and DR, see Refs.~\cite{Buen-Abad:2015ova, Lesgourgues:2015wza, Tang:2016mot} for detailed discussions. 
Here we provide an intuitive picture to show why the case with gauge symmetry breaking into $U(1)$ 
subgroup does not have the required properties. The Feynman diagrams for DM-DR scattering within $U(1)$ are 
shown in Fig.~\ref{fig:u1}, where $A^m$ is the massive vector DM and $A^a$ is the massless $U(1)$ DR. 
When the temperature $T$ is below the DM mass, we can immediately estimate that the cross section should 
have the following form,   $\sigma \sim g^4/m_{A^m}^2$,  similar to the Thomson scattering where $m_{A^m}$ 
is replaced by the electron's mass. If the DM's mass $m_{A^a}$ is around the electroweak scale, the cross section 
and scattering rate would be too small to have effects in the late-time universe. Therefore, such a scenario can not 
change $\sigma_8$ and would not be able to resolve the tension. However, in the non-abelian case, we have 
diagrams like Fig.~\ref{fig:scattering}(left) where the vertex with triple massless gauge bosons shows up. 
This leads to the DM-DR scattering cross section scaling as $\sigma \sim g^4/T^2$,
which could grow large as the universe cools down. So the scattering rate could actually be important 
in the whole radiation-dominated era and the structure growth at the corresponding scales can be affected. 

Also, if the gauge group is completely broken, there will be no DR left in the model, although massive bosons are still DM candidates~\cite{Hambye:2008bq, Baek:2012se,Farzan:2012hh, Carone:2013wla, Hambye:2013sna, Chiang:2013kqa, Baek:2014goa, Chen:2015dea, DiChiara:2015bua, Gross:2015cwa, Karam:2016rsz}. 

\section{Phenomenology and Constraints}\label{sec:constraint}
Phenomenologies of the above model in particle physics, cosmology and astrophysics can be very rich, depending on the mass spectra and the coupling strengths. There could be confinement if gauge coupling $g$ is large. Neglecting mass threshold effect, the confinement scale $\Lambda$ for residual $SU(2)$ is obtained by 
\[
\Lambda = \mu_0\exp \left[-1\bigg/\dfrac{g^2\left(\mu_0\right)}{8\pi^2} \left(\dfrac{11}{3}\times 2 - \dfrac{1}{6}\right)\right],
\]
where $\mu_0$ is some reference scale at which $g(\mu_0)$ is given. Below confinement scale, we shall expect  that hidden glueball associated with unbroken gauge symmetry with mass $\sim \Lambda$ could also be dark matter candidate. Phenomenology about glueball DM has been well considered, see Refs.~\cite{Forestell:2016qhc, Kribs:2016cew, Soni:2016gzf, Yamanaka:2014pva, Boddy:2014yra, Faraggi:2000pv}. Moreover, depending on the relative size of confinement scale and $v_\phi$, the vectorial bound states of $\Phi$, other than glueball or $A^{4,...,8}$, might be the DM candidates~\cite{Hambye:2009fg}. In the rest of our paper, we shall not pursue the confinement case further.

Now we focus on the unconfined case with relatively small $g$. For example, we check that if $g(\mu_0=1\TeV)\sim 10^{-1}$, $\Lambda \sim 10^{-500}\TeV$ for the residual $SU(2)$ and is much smaller than the current temperature of our Universe, $T_\gamma \sim 2.73\textrm{K} \sim 8.6\times 10^{-5} \eV$. In such a case, there will be no hidden glueballs or other bound states but only massless self-interacting gluon radiation in cosmic background. 

\begin{figure}[t]
	\includegraphics[scale=0.6]{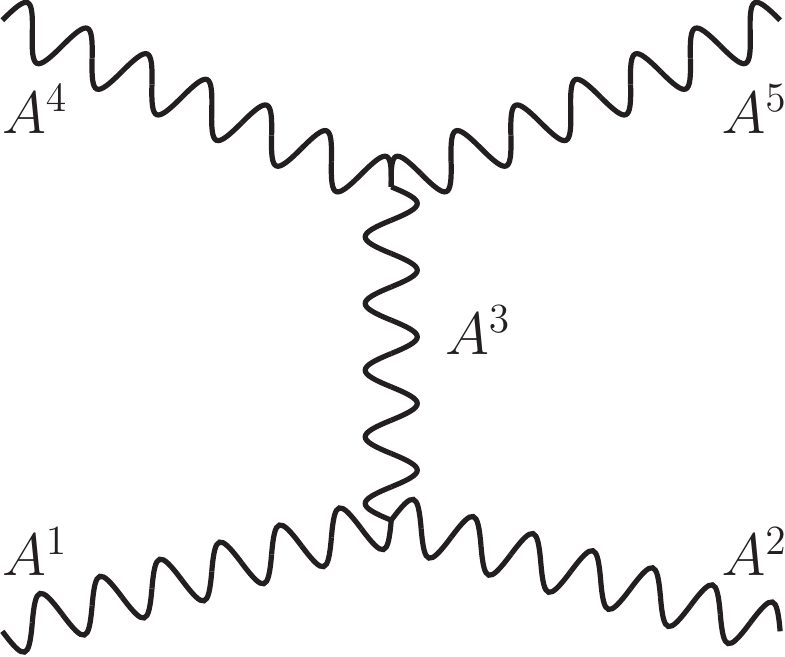} \qquad \qquad
	\includegraphics[scale=0.6]{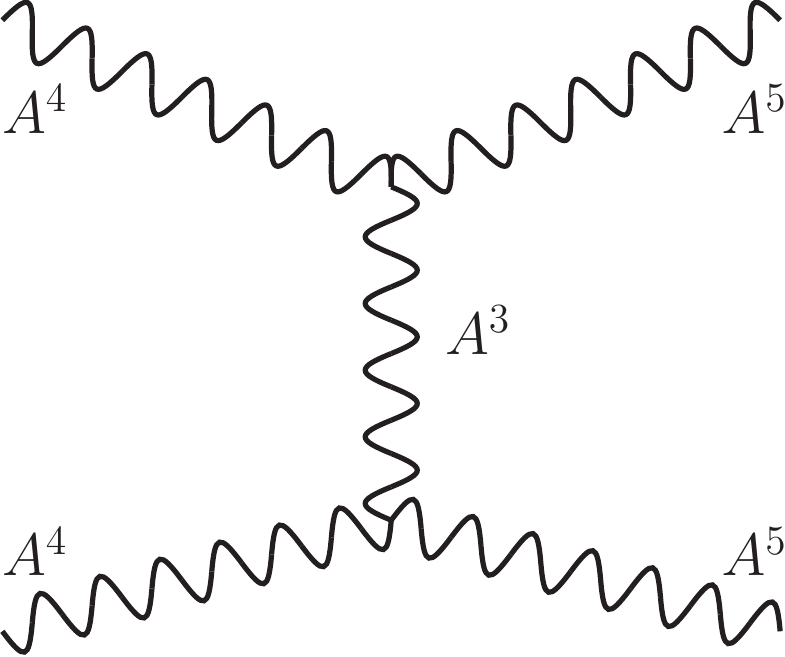}
	\caption{Typical Feynmann diagrams for DM-DR (left) and DM-DM (right) scattering.\label{fig:scattering}}
\end{figure}

{\it DM--DR Scattering}: 
Suppose we have massless DR $A^a$ with temperature $T_A$ in cosmic background. 
Then VDM $A^m$ should scatter with this DR through such a diagram shown in 
Fig.~\ref{fig:scattering} (left). This may induce collisional damping in the matter power spectrum~\cite{Boehm:2004th, Green:2005fa, Loeb:2005pm}, which would be similar to baryonic acoustic oscillation with different scale-dependence. The gauge coupling that is still allowed to have sizable effects on LSS is related with DM masses through~\cite{Ko:2016uft}
\begin{equation}
g^2 \lesssim \frac{T_\gamma}{T_A}\left(\frac{m_A}{M_{P}}\right)^{1/2},
\end{equation}
where the Planck scale $M_{P}\sim 10^{18} \GeV$. For $T_A\sim T_\gamma$ and $m_A\sim 10\TeV $, we have $g\sim 10^{-3.5}$. The suppression effect leads to a smaller $\sigma_8$, which can help to relax the tension between {\tt Planck}'s data and other experimental results~\cite{Heymans:2012gg}. With such a small coupling, VDM $A^{m}$ cannot be the usual 
thermal dark matter and we shall provide an alternative production mechanism with low reheating temperature in Sec.~\ref{sec:relic}. 

{\it Self-Interaction}: The residual massless gauge bosons $A^a$ may mediate 
large self-scattering cross section for VDM $A^m$ if the gauge coupling 
is large, see Fig.~\ref{fig:scattering} (right). Because $A^{8}$ only interacts with massive 
$A^{4,5,6,7}$, its self-scattering is much smaller and can be neglected. Large DM self-interaction can change the dark halo's shape and density profile, and observation of the offset between DM and gas in Bullet Cluster collision. The most stringent constraint is  from galactic dynamics~\cite{Carroll:mha}, which should be imposed on the parameter $\sigma_{AA}/m_{A}$, 
where $\sigma_{AA}$ is the cross section for DM $A+A\rightarrow A+A$ elastic scattering 
process. For $m_{A}\sim 10\TeV$, $g$ should be less than $1$. As we have shown above, 
this constraint can be easily satisfied for our interested region where $g\sim 10^{-3.5}$.

{\it Dark Radiation}: The massless gauge bosons $A^a$s associated with the unbroken $SU(N-1)$ potentially can contribute considerably to the energy density of radiation in cosmic background, namely shifting the effective number of neutrinos, $N_\eff$. The production of massless $A^a$ is mainly through the decay of $\phi$, because the thermal production is too small due to the smallness of the loop-induced coupling. The decay width of $\phi\rightarrow A^a + A^a$ from the $A^m$-loop can be estimated as
\begin{equation}\label{eq:decaywidth}
\Gamma_A \sim \frac{g^4 m^3_\phi}{\left(4\pi\right)^5 v_\phi^2}\sim \frac{g^6 m^3_\phi}{\left(4\pi\right)^5 m^2_A}.
\end{equation}
One may introduce additional new heavy $SU(N)$ charged fermion/scalar to increase the decay width. 
%As long as $\phi$ decays before CMB temperature drop around $1\eV$, the resulting DR would be able to scatter with DM.

One subtlety is that $\phi$ might decay non-relativistically. In such a case, considerable entropy will be produced, which could result in too large $\delta N_\eff$. This problem can be easily evaded once we allow a Higgs portal coupling $\lambda_{\phi H} \Phi^\dagger \Phi H^\dagger H$~\cite{Baek:2013qwa} ($H$ is the SM Higgs doublet), then $\phi$'s would also ``heat up" the SM thermal bath before BBN. By tuning the size of $\lambda_{\phi H}$ and $\phi$'s decay branch ratios, we can get any required $\delta N_\eff$. 

\begin{figure}[t]
	\includegraphics[width=0.5\textwidth,height=0.4\textwidth]{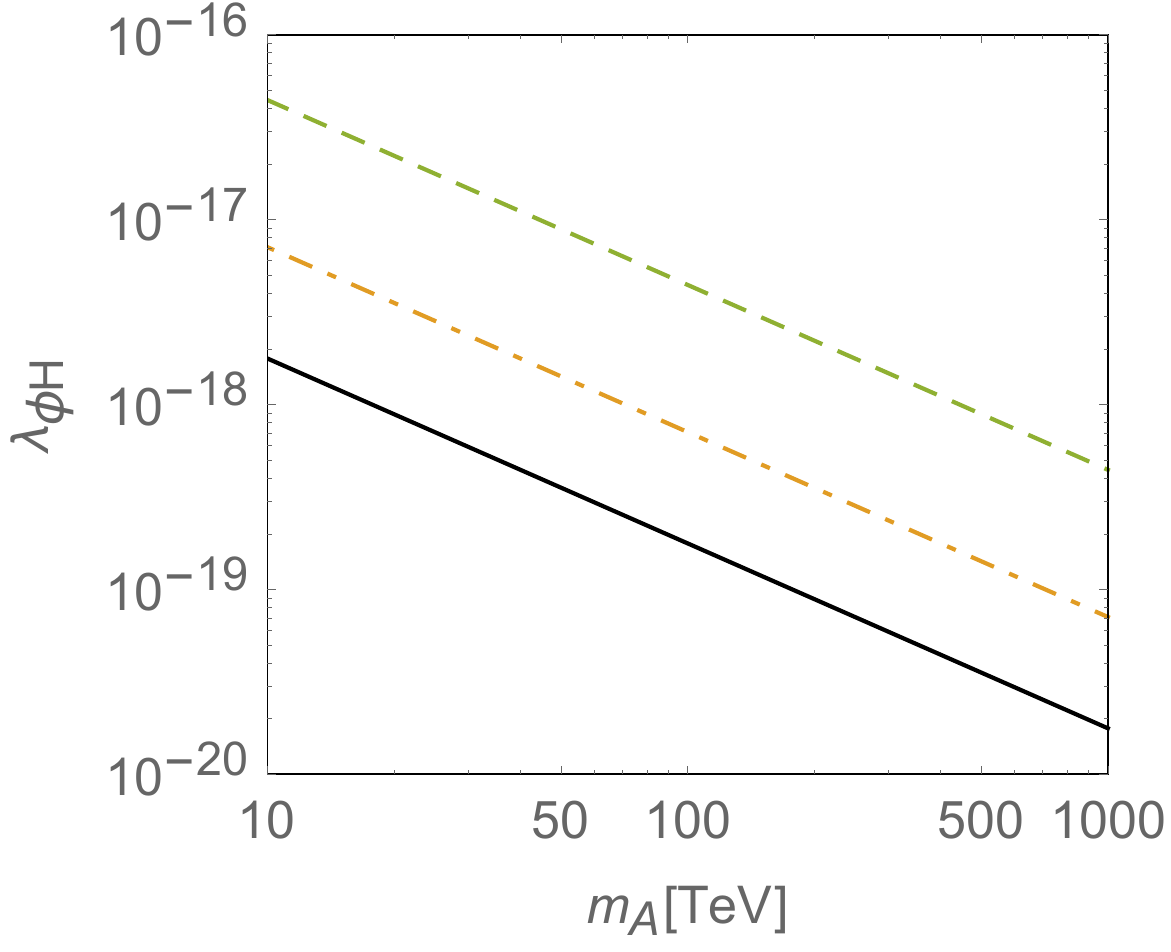}
	\caption{$\lambda_{\phi H}$ as function of DM mass $m_A$. The solid, dot-dashed and dashed lines correspond to $m_\phi=1,2,5\TeV$, respectively.}
	\label{fig:lambda}
\end{figure}

With the Higgs portal coupling, $\phi$ can decay into two SM Higgs bosons, $h$, which quickly decay into other SM particles. The decay width for $\phi\rightarrow h h$ is around 
\begin{equation}
\Gamma_{h}\sim \frac{\lambda^2_{\phi H}v^2_\phi}{16\pi^2 m_\phi}.
\end{equation}
With $\Gamma_A/\Gamma_{h}\sim 0.1$ to have considerable DR $A^a$ as $\delta N_\eff\sim 0.6$, we can estimate 
\begin{equation}\label{eq:portal}
\frac{1}{\left(4\pi\right)^3 }\frac{g^4m^4_\phi}{\lambda^2_{\phi H}v^4_\phi}\sim 0.1.
\end{equation}
If we take $g\sim 10^{-3.5}, v_\phi\sim 10^{4.5}\TeV$ and $m_\phi\sim 1\TeV$, we would get $\lambda_{\phi H}\sim 10^{-18}$. We show in Fig.~\ref{fig:lambda} how $\lambda_{\phi H}$ changes as function of DM mass $m_A$ for $m_{\phi}=1,2,5\TeV$. The smallness of $\lambda_{\phi H}$ also indicates that the dark sector can not be in thermal equilibrium with SM.

Current limit on $\delta N_\eff$ from {\tt Planck}~\cite{Planck:2015xua} in standard $\Lambda$CDM cosmology 
with six cosmological parameters gives $\delta N_\eff \lesssim 0.7$ with $95\%$ confidence level. 
Future prospect with $\delta N_\eff\lesssim 0.1$ would provide much more powerful limit. 
The upper limit is relaxed if $N_\eff$ is treated as an additional cosmological parameter.   Note that $N_\eff\simeq 
0.4-1$ may help to resolve the conflict between HST data and {\tt Planck}'s results~\cite{Riess:2016jrr}.

{\it Direct and Indirect Detection}: Since VDM has mass around $\sim 10$ 
TeV and very small coupling $g\sim 10^{-3}$, bounds from current DM direct, indirect and collider experiments are all evaded, which also means that it would be difficult to test this model by other means other than cosmological and astrophysical effects mentioned above. This picture can be changed if there is a non-renormalizable operator, a dim-6 kinetic mixing between $A^8$ and $U(1)_Y$ field strength $B_{\mu\nu}$ through $\Phi^\dagger F^{\mu\nu} \Phi B_{\mu\nu} / M^2$. This operator will induce the effective kinetic mixing $\kappa$ at the level of 
$\kappa \sim v_\phi^2 / M^2$ so that the VDM $A^8$ can decay and leave detectable signatures in cosmic rays, gamma rays and neutrino flux if $M\sim 10^{18}\GeV$.

\section{Relic Abundance and Thermal History}\label{sec:relic}
We investigate the production of massive VDM $A^m$ through freeze-in mechanism by $\phi  +\phi \rightarrow  A^{m} + A^{m}$. The cross section for this process behaves as $g^4/E^2$, where $E$ is total energy of the colliding $\phi$s. 
When $m_{A} \gg T$ ($T$ is thermal temperature of $\phi$, which could be different from the one shared by SM thermal bath), effectively only those $\phi$'s with $E\gg T$ at the high energy tail of thermal distribution are energetic enough to produce massive $A^m$. Therefore, we expect there should be a Boltzmann suppression factor like $\exp\left(-E/T\right)$. We can solve the Boltzmann equation
\begin{equation}
\frac{dn_A}{dt}+3\mathcal{H}n_A = \int d\Pi_1 d\Pi_2 d\Pi_3 d\Pi_4 (2\pi)^4\delta^4\left(p_1+p_2-p_3-p_4\right)|\mathcal{M}|^2 f_1 f_2,
\end{equation}
where $d\Pi_i\equiv d^3p_i/(2\pi)^3$, $\mathcal{M}$ is the matrix element for $\phi +\phi \rightarrow A^{m} +A^{m}$, and $f_i$ is the Bose-Einstein distribution, $f_i=1/(\exp[E_i/T]-1)\simeq \exp[-E_i/T]$. We have neglected the reverse process due to the smallness of number of $A^{m}$s. Rewrite the right-handed side approximately as
\begin{equation}\label{eq:collision}
\int d\Pi_1 d\Pi_2  \exp\left(-\frac{E_1+E_2}{T}\right)\frac{g^4}{\left(E_1+E_2\right)^2},
\end{equation}
where the integration is over $E_1+E_2\geq 2m_{A}$, then we can follow the standard 
procedure to do the numerical computation as Ref.~\cite{Gondolo:1990dk}. 
In the following, we will instead  try to estimate the size of $g$ analytically by making some 
simplifications. We approximate Eq.~(\ref{eq:collision}) with 
\begin{equation}
\frac{g^4}{T^2}\exp\left[-\frac{m_A}{T}\right]T^6,
\end{equation}
which has the correct dimension and should be able to give conservative estimation. Introducing $Y_A\equiv n_A/s$, where $s\sim T^3$ is the entropy density, we have $dn_A/dt+3\mathcal{H}n_A =sdY_A/dt$ and 
\begin{equation}
\frac{dY_A}{dt}=\frac{g^4}{s}\exp\left[-\frac{m_A}{T}\right]T^4.
\end{equation}
In the radiation-dominant epoch, we have $dt=-dT/(\mathcal{H}T)$ and can rewrite the Boltzmann equation as
\begin{equation}
\frac{dY_A}{dT}=-\frac{g^4 }{s \mathcal{H}T}\exp\left[-\frac{m_A}{T}\right]T^4\simeq -\frac{g^4 M_{P} }{T^2}\exp\left[-\frac{m_A}{T}\right].
\end{equation}
We can solve the above differential equation and get 
\begin{equation}
Y_A\simeq \frac{g^4 M_{P} }{m_A}\exp\left[-\frac{m_A}{T_{\textrm{reh}}}\right],
\end{equation}
where $T_{\textrm{reh}}$ is the reheating temperature after inflation. 
To get the correct relic abundance for VDM $A^m$, we should have
\begin{equation}
Y_A=\frac{\Omega_X m_p}{\Omega_b m_A}\eta,
\end{equation}
where $\Omega_b$ and $\Omega_A$ are the energy density fractions of baryon and dark matter, respectively, $\Omega_X/\Omega_b\simeq 5$, $m_p\simeq 1\GeV$ is proton mass and $\eta\simeq 6\times 10^{-10}$ is baryon-to-photon ratio. Finally, we have the relation for $m_A/T_{\textrm{reh}}$,
\begin{equation}\label{eq:gaugecpl}
\frac{m_A}{T_{\textrm{reh}}}\sim \ln \left[\frac{\Omega_b M_P g^4}{\Omega_X m_p \eta} \right]. 
\end{equation}
If $g^2\simeq 10^{-7}$, which is also large enough for thermalizing massless $A^a$, we can get $m_A/T_{\textrm{reh}}\sim 30$. It is also expected that larger $g^2$ would require larger $m_A/T_{\textrm{reh}}$ or relatively smaller $T_{\textrm{reh}}$. 

In the above rough estimation, we have assumed  instantaneous reheating for simplicity. 
The effects from reheating over a finite period time are surely important, but is more model-dependent. 
One immediate effect is from the fact that the highest temperature $T_{\textrm{high}}$ might be larger than 
the usual reheating temperature $T_\textrm{reh}$~\cite{Giudice:2000ex}, then we should substitute 
$T_\textrm{reh}$ with $T_{\textrm{high}}$ in the above calculation and also consider the corresponding 
entropy production factor. 
	
We also need to consider the dilution effect from entropy production due to the late-time decay of $\phi$. 
In the simultaneous decay approximation, the ratio for the entropy before and after the $\phi$ decay is given by~\cite{Scherrer:1984fd} 
\begin{equation}
	\frac{S_\textrm{after}}{S_\textrm{before}}\sim \frac{m_\phi}{\sqrt{\Gamma M_P}},
\end{equation}
where $\Gamma=\Gamma_A+\Gamma_h$ is the decay width of $\phi$. If we take $g\sim 10^{-3.5}, v_\phi\sim 10^{4.5}\TeV$ and $m_\phi\sim 1\TeV$, we have $S_\textrm{after}/S_\textrm{before}\sim 10^6$. Then with the dilution factor, we actually need a smaller $m_A/T_\textrm{reh}\sim 18$ to get the right relic density. 

\begin{figure}[t]
	\includegraphics[width=0.5\textwidth,height=0.4\textwidth]{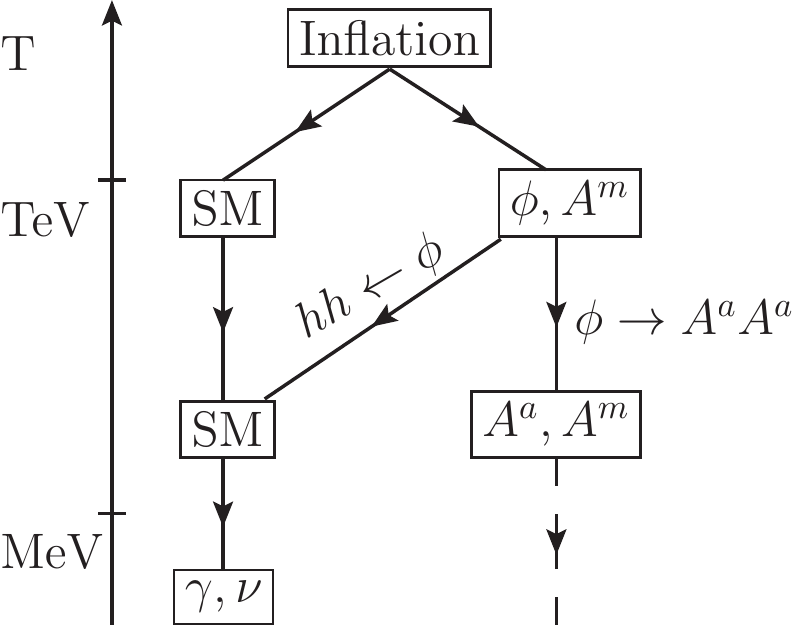}
	\caption{Schematic picture for thermal history of DM $A^m$, DR $A^a$, dark Higgs boson $\phi$ and SM.}
	\label{fig:hist}
\end{figure}

{\it Thermal History:} Now we are in a position to discuss the full thermal history of our model. If we restrict 
to the minimal model setup without introducing other new fields or interactions except inflation, then the overall 
picture of thermal history in our scenario is shown schematically in Fig.~\ref{fig:hist}. After inflation and reheating, dark sector and SM sector have already decoupled due to the tiny $\lambda_{\phi H} \sim 10^{-18}$.  
As shown above, DM $A^m$ is mostly produced at the high temperature due to the freeze-in mechanism, 
$\phi + \phi \rightarrow A^m + A^m$. This is because $A^m$ only has gauge interactions which is not large 
enough to make it as thermal DM. Since $\phi$ couples to DR $A^a$ at one-loop level as Fig.~\ref{fig:decay}, 
its decay width is small so that its lifetime becomes long. Therefore, $\phi$ decays non-relativistically into both 
DR $A^a$ and the SM sector. In the decay process, entropy is produced and we can adjust the branch ratios 
with proper $\lambda_{\phi H}$ as Eq.~\ref{eq:portal} to get sizable $\delta N_\eff$ for DR $A^a$.

\section{Numeric Results}\label{sec:numeric}

To visualize DM-DR's effects on matter power spectrum, we have modified the public code {\tt Class}~\cite{class} to implement the coupled Boltzmann equations~\cite{Ma:1995ey},
\begin{align}
\dot{\theta}_m &= k^2\Psi -\mathcal{H} \theta_m+ S^{-1}\dot{\mu}\left(\theta_r-\theta_{m}\right),\label{eq:veldiv1}\\
\dot{\theta}_r &= k^2 \Psi +k^2\left(\delta_r/4 -\sigma_r\right)-\dot{\mu}\left(\theta_r-\theta_{m}\right),\label{eq:veldiv2}
\end{align}
where dot means derivative over conformal time $d\tau\equiv dt/a$ ($a$ is the scale factor),  
$\theta_r$ and $\theta_m$ are velocity divergences of DR $A^a$ and DM $A^m$, 
$k$ is the co-moving wavenumber, $\Psi$ is the gravitational potential, $\delta_r$ and 
$\sigma_r$ ($\sigma_r=0$ for $A^a$ as perfect fluid) are the density perturbation and the anisotropic stress potential of $A^a$, and 
$\mathcal{H} \equiv \dot{a}/a$ is the conformal Hubble parameter. The interacting rate of DM-DR scattering and the energy density ratio are defined by $\dot{\mu}=a n_{A^m}\langle\sigma_{A^m A^a}c\rangle$ and $S=3\rho_{A^m}/4\rho_{A^a}$, respectively. 

\begin{figure}[t]
	\includegraphics[width=0.45\textwidth,height=0.42\textwidth]{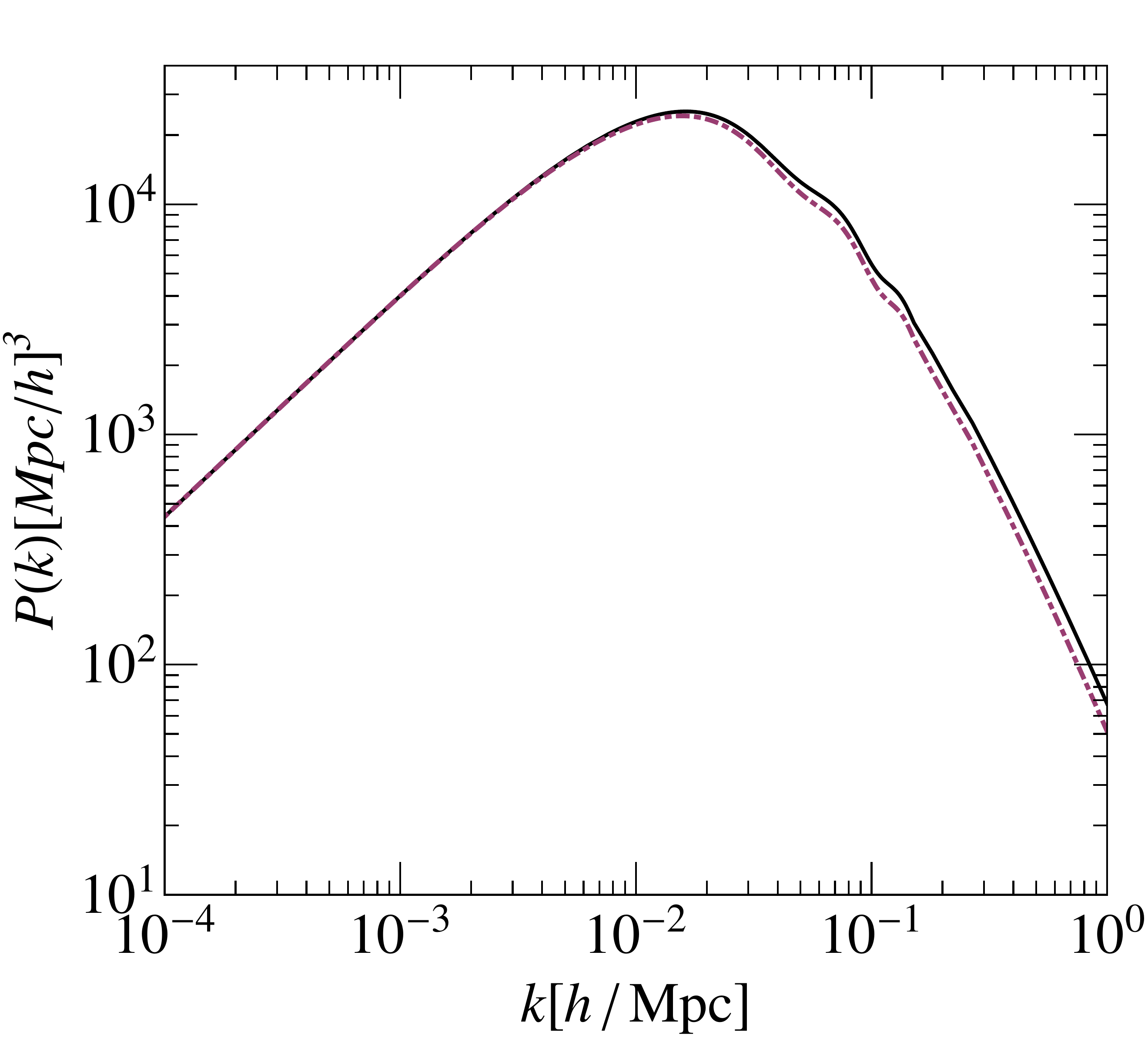}
	\includegraphics[width=0.45\textwidth,height=0.42\textwidth]{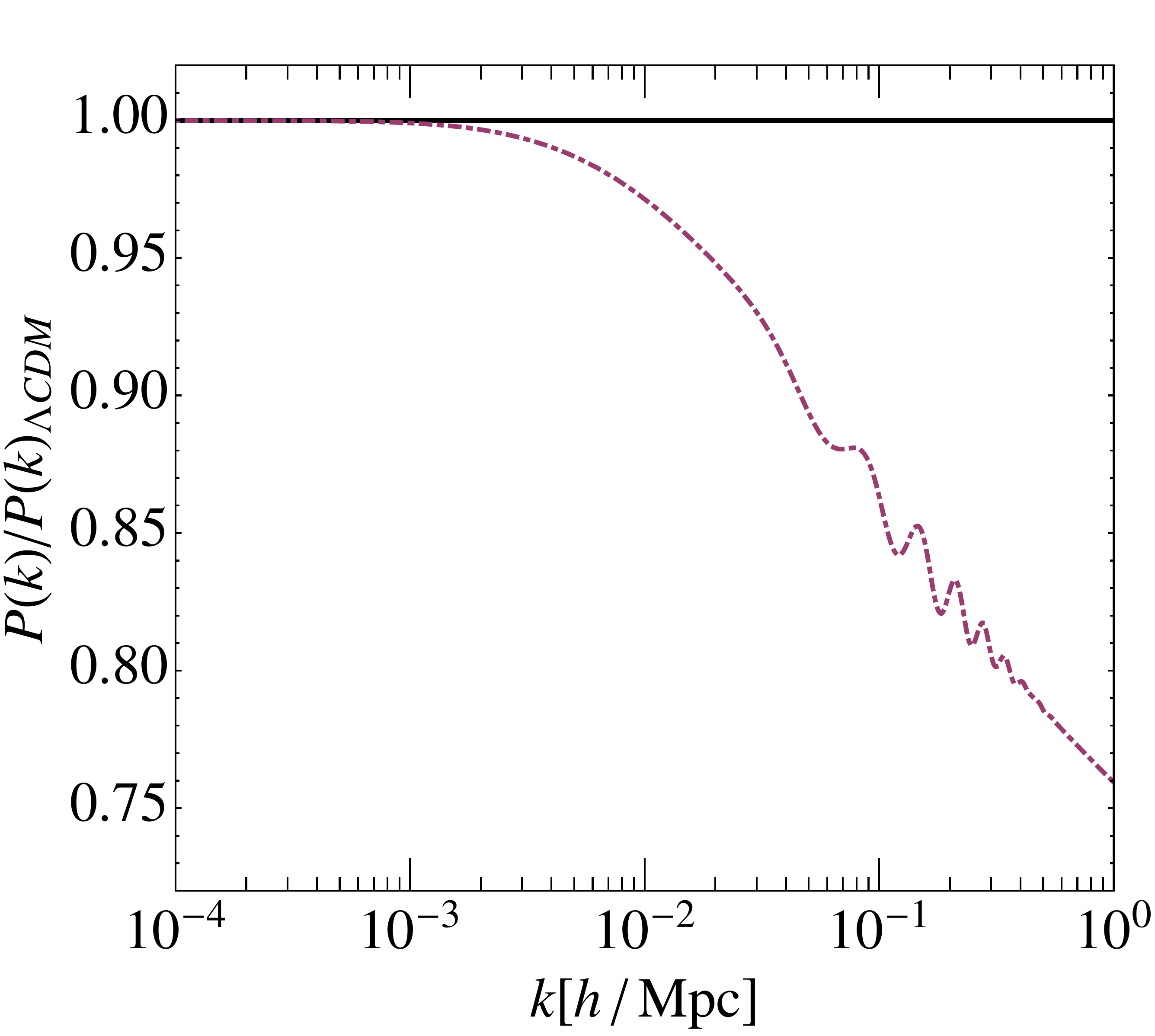}
	\caption{Matter power spectrum $P(k)$ (left) and ratio (right) with $m_\chi\simeq 10\TeV$ and $g^2_X\simeq 10^{-7}$, in comparison with $\Lambda$CDM. The black solid lines are for $\Lambda$CDM and the purple dot-dashed lines for interacting DM-DR case, with input parameters in Eq.~\ref{eq:input}. We can easily see that $P(k)$ is suppressed for modes that enter horizon at radiation-dominant era. Those little wiggles are due to the well-known baryon acoustic oscillation.}
	\label{fig:sigma8}
\end{figure}

We illustrate the effects on matter power spectrum $P(k)$ in Fig.~\ref{fig:sigma8} where the solid lines are the outputs from $\Lambda$CDM and the dot-dashed lines are for interacting DM-DR scenario. The left panel shows the overall picture of $P(k)$ which the right one shows the ratio. It is clearly that power spectrum can be suppressed when DM-DR interaction is considered. We have taken the central values of six parameters of $\Lambda$CDM from $\planck$~\cite{Planck:2015xua},
\begin{align}\label{eq:input}
&\Omega_b h^2=0.02227, \Omega _c h^2 =0.1184, 100\theta_{\textrm{MC}}=1.04106, \nonumber \\
&\tau = 0.067, \ln\left(10^{10}A_s\right)=3.064, n_s=0.9681,
\end{align} 
and treat neutrino mass the same way as $\planck$ did with $\sum m_\nu=0.06\eV$, which gives $\sigma_8=0.815$ in vanilla $\Lambda$CDM cosmology. Together with the same inputs as above, we take $\delta N_\eff \simeq 0.5$,  $m_\chi\simeq 10\TeV$ and $g_X^2\simeq 10^{-7}$ in the interacting DM-DR case, we have $\sigma_8\simeq 0.746$ which is much closer to the value $\sigma_8\simeq 0.730$ given by weak lensing survey CFHTLenS~\cite{Heymans:2012gg}.

\section{Conclusion}\label{sec:conclusion}
In this paper, we have proposed a particle physics model in which vector dark matter (VDM) and 
dark radiation (DR) have a common origin, namely a Yang-Mills dark sector. We have explicitly 
shown an illustrating case where dark $SU(3)$ gauge group is spontaneously broken to its $SU(2)$
subgroup. The residual massless gauge bosons constitute DR while other massive ones make 
up the VDM. Interestingly, VDM naturally can scatter with DR through the original Yang-Mills self-interaction,  which can lead to a suppressed matter power spectrum and give rise to a smaller $\sigma_8$ for structure growth. On the other hand, DR also helps to resolve the tension in Hubble constant measurements between {\tt Planck}~\cite{Planck:2015xua} and Hubble Space Telescope~\cite{Riess:2016jrr}.

In the minimal model with one fundamental scalar that breaks the dark gauge symmetry, 
the new massive Higgs boson only couples to residual massless gauge boson at loop level, 
so the coupling would be too small to thermalize DR. However, production of DR could be due to the decay of this new Higgs. The viable way to produce DM with correct relic abundance can be realized in freeze-in mechanism, provided the reheating temperature is about one order-of-magnitude smaller than DM mass. 

\begin{acknowledgments}
This work is supported in part by National Research Foundation of Korea 
(NRF) Research Grant NRF-2015R1A2A1A05001869 (PK,YT), and by the NRF grant 
funded by the Korea government (MSIP) (No. 2009-0083526) through Korea 
Neutrino Research Center at Seoul National University (PK). YT is also supported by the Grant-in-Aid for Innovative Areas No.16H06490 in Japan.
\end{acknowledgments}

%\bibliographystyle{../utphysMa}
%\bibliography{../references}

\begin{thebibliography}{10}
	
	\bibitem{Bringmann:2013vra}
	T.~Bringmann, J.~Hasenkamp, and J.~Kersten, {\it {Tight bonds between sterile
			neutrinos and dark matter}},
	\href{http://dx.doi.org/10.1088/1475-7516/2014/07/042}{{\em JCAP} {\bfseries
			1407} (2014) 042} [\href{http://arxiv.org/abs/1312.4947}{{\ttfamily
			arXiv:1312.4947}}].
	%%CITATION = ARXIV:1312.4947;%%.
	
	\bibitem{Boehm:2014vja}
	C.~Boehm, J.~A. Schewtschenko, R.~J. Wilkinson, C.~M. Baugh, and S.~Pascoli,
	{\it {Using the Milky Way satellites to study interactions between cold dark
			matter and radiation}},
	\href{http://dx.doi.org/10.1093/mnrasl/slu115}{{\em Mon. Not. Roy. Astron.
			Soc.} {\bfseries 445} (2014) L31--L35}
	[\href{http://arxiv.org/abs/1404.7012}{{\ttfamily arXiv:1404.7012}}].
	%%CITATION = ARXIV:1404.7012;%%.
	
	\bibitem{Ko:2014bka}
	P.~Ko and Y.~Tang, {\it {$\nu \Lambda$MDM: A Model for Sterile Neutrino and
			Dark Matter Reconciles Cosmological and Neutrino Oscillation Data after
			BICEP2}},
	\href{http://dx.doi.org/10.1016/j.physletb.2014.10.035}{{\em Phys.Lett.}
		{\bfseries B739} (2014) 62--67}
	[\href{http://arxiv.org/abs/1404.0236}{{\ttfamily arXiv:1404.0236}}].
	%%CITATION = ARXIV:1404.0236;%%.
	
	\bibitem{Chu:2014lja}
	X.~Chu and B.~Dasgupta, {\it {Dark Radiation Alleviates Problems with Dark
			Matter Halos}},
	\href{http://dx.doi.org/10.1103/PhysRevLett.113.161301}{{\em Phys. Rev. Lett.}
		{\bfseries 113} no.~16, (2014) 161301}
	[\href{http://arxiv.org/abs/1404.6127}{{\ttfamily arXiv:1404.6127}}].
	%%CITATION = ARXIV:1404.6127;%%.
	
	\bibitem{Buen-Abad:2015ova}
	M.~A. Buen-Abad, G.~Marques-Tavares, and M.~Schmaltz, {\it {Non-Abelian dark
			matter and dark radiation}},
	\href{http://dx.doi.org/10.1103/PhysRevD.92.023531}{{\em Phys. Rev.} {\bfseries
			D92} no.~2, (2015) 023531} [\href{http://arxiv.org/abs/1505.03542}{{\ttfamily
			arXiv:1505.03542}}].
	%%CITATION = ARXIV:1505.03542;%%.
	
	\bibitem{Lesgourgues:2015wza}
	J.~Lesgourgues, G.~Marques-Tavares, and M.~Schmaltz, {\it {Evidence for dark
			matter interactions in cosmological precision data?}},
	\href{http://dx.doi.org/10.1088/1475-7516/2016/02/037}{{\em JCAP} {\bfseries
			1602} no.~02, (2016) 037} [\href{http://arxiv.org/abs/1507.04351}{{\ttfamily
			arXiv:1507.04351}}].
	%%CITATION = ARXIV:1507.04351;%%.
	
	\bibitem{Tang:2016mot}
	Y.~Tang, {\it {Interacting Scalar Radiation and Dark Matter in Cosmology}},
	\href{http://dx.doi.org/10.1016/j.physletb.2016.04.026}{{\em Phys. Lett.}
		{\bfseries B757} (2016) 387--392}
	[\href{http://arxiv.org/abs/1603.00165}{{\ttfamily arXiv:1603.00165}}].
	%%CITATION = ARXIV:1603.00165;%%.
	
	\bibitem{Bringmann:2016ilk}
	T.~Bringmann, H.~T. Ihle, J.~Kersten, and P.~Walia,
	{\it {Suppressing structure formation at dwarf galaxy scales and below: late
			kinetic decoupling as a compelling alternative to warm dark matter}},
	[\href{http://arxiv.org/abs/1603.04884}{{\ttfamily arXiv:1603.04884}}].
	%%CITATION = ARXIV:1603.04884;%%.
	
	\bibitem{Ko:2016uft}
	P.~Ko and Y.~Tang,
	{\it {Light dark photon and fermionic dark radiation for the Hubble constant
			and the structure formation}},
	[\href{http://arxiv.org/abs/1608.01083}{{\ttfamily arXiv:1608.01083}}].
	%%CITATION = ARXIV:1608.01083;%%.
	
	\bibitem{Planck:2015xua}
	{\bfseries Planck Collaboration} , P.~Ade {\em et al.},
	{\it {Planck 2015 results. XIII. Cosmological parameters}},
	[\href{http://arxiv.org/abs/1502.01589}{{\ttfamily arXiv:1502.01589}}].
	%%CITATION = ARXIV:1502.01589;%%.
	
	\bibitem{Riess:2016jrr}
	A.~G. Riess {\em et al.}, {\it {A 2.4\% Determination of the Local Value of the
			Hubble Constant}},
	\href{http://dx.doi.org/10.3847/0004-637X/826/1/56}{{\em Astrophys. J.}
		{\bfseries 826} no.~1, (2016) 56}
	[\href{http://arxiv.org/abs/1604.01424}{{\ttfamily arXiv:1604.01424}}].
	%%CITATION = ARXIV:1604.01424;%%.
	
	\bibitem{Heymans:2012gg}
	C.~Heymans {\em et al.}, {\it {CFHTLenS: The Canada-France-Hawaii Telescope
			Lensing Survey}},
	\href{http://dx.doi.org/10.1111/j.1365-2966.2012.21952.x}{{\em Mon. Not. Roy.
			Astron. Soc.} {\bfseries 427} (2012) 146}
	[\href{http://arxiv.org/abs/1210.0032}{{\ttfamily arXiv:1210.0032}}].
	%%CITATION = ARXIV:1210.0032;%%.
	
	\bibitem{Pourtsidou:2016ico}
	A.~Pourtsidou and T.~Tram,
	{\it {Reconciling CMB and structure growth measurements with dark energy
			interactions}},  [\href{http://arxiv.org/abs/1604.04222}{{\ttfamily
			arXiv:1604.04222}}].
	%%CITATION = ARXIV:1604.04222;%%.
	
	\bibitem{DiValentino:2016hlg}
	E.~Di~Valentino, A.~Melchiorri, and J.~Silk,
	{\it {Reconciling Planck with the local value of $H_0$ in extended parameter
			space}},  [\href{http://arxiv.org/abs/1606.00634}{{\ttfamily
			arXiv:1606.00634}}].
	%%CITATION = ARXIV:1606.00634;%%.
	
	\bibitem{Qing-Guo:2016ykt}
	Q.-G. Huang and K.~Wang,
	{\it {How the Dark Energy Can Reconcile \textit{Planck} with Local
			Determination of the Hubble Constant}},
	[\href{http://arxiv.org/abs/1606.05965}{{\ttfamily arXiv:1606.05965}}].
	%%CITATION = ARXIV:1606.05965;%%.
	
	\bibitem{Archidiacono:2016kkh}
	M.~Archidiacono, S.~Gariazzo, C.~Giunti, S.~Hannestad, R.~Hansen, M.~Laveder,
	and T.~Tram,
	{\it {Pseudoscalar - sterile neutrino interactions: reconciling the cosmos with
			neutrino oscillations}},  [\href{http://arxiv.org/abs/1606.07673}{{\ttfamily
			arXiv:1606.07673}}].
	%%CITATION = ARXIV:1606.07673;%%.
	
	\bibitem{Wyman:2013lza}
	M.~Wyman, D.~H. Rudd, R.~A. Vanderveld, and W.~Hu, {\it {Neutrinos Help
			Reconcile Planck Measurements with the Local Universe}},
	\href{http://dx.doi.org/10.1103/PhysRevLett.112.051302}{{\em Phys. Rev. Lett.}
		{\bfseries 112} no.~5, (2014) 051302}
	[\href{http://arxiv.org/abs/1307.7715}{{\ttfamily arXiv:1307.7715}}].
	%%CITATION = ARXIV:1307.7715;%%.
	
	\bibitem{Zhang:2014lfa}
	J.-F. Zhang, Y.-H. Li, and X.~Zhang, {\it {Measuring growth index in a universe
			with sterile neutrinos}},
	\href{http://dx.doi.org/10.1016/j.physletb.2014.10.044}{{\em Phys. Lett.}
		{\bfseries B739} (2014) 102--105}
	[\href{http://arxiv.org/abs/1408.4603}{{\ttfamily arXiv:1408.4603}}].
	%%CITATION = ARXIV:1408.4603;%%.
	
	\bibitem{DiValentino:2016ucb}
	E.~Di~Valentino and F.~R. Bouchet,
	{\it {A comment on power-law inflation with a dark radiation component}},
	[\href{http://arxiv.org/abs/1609.00328}{{\ttfamily arXiv:1609.00328}}].
	%%CITATION = ARXIV:1609.00328;%%.
	
	\bibitem{Baek:2013dwa}
	S.~Baek, P.~Ko, and W.-I. Park, {\it {Hidden sector monopole, vector dark
			matter and dark radiation with Higgs portal}},
	\href{http://dx.doi.org/10.1088/1475-7516/2014/10/067}{{\em JCAP} {\bfseries
			1410} no.~10, (2014) 067} [\href{http://arxiv.org/abs/1311.1035}{{\ttfamily
			arXiv:1311.1035}}].
	%%CITATION = ARXIV:1311.1035;%%.
	
	\bibitem{Khoze:2014woa}
	V.~V. Khoze and G.~Ro, {\it {Dark matter monopoles, vectors and photons}},
	\href{http://dx.doi.org/10.1007/JHEP10(2014)061}{{\em JHEP} {\bfseries 10}
		(2014) 61} [\href{http://arxiv.org/abs/1406.2291}{{\ttfamily
			arXiv:1406.2291}}].
	%%CITATION = ARXIV:1406.2291;%%.
	
	\bibitem{Kawasaki:2015lpf}
	M.~Kawasaki, F.~Takahashi, and M.~Yamada, {\it {Suppressing the QCD Axion
			Abundance by Hidden Monopoles}},
	\href{http://dx.doi.org/10.1016/j.physletb.2015.12.075}{{\em Phys. Lett.}
		{\bfseries B753} (2016) 677--681}
	[\href{http://arxiv.org/abs/1511.05030}{{\ttfamily arXiv:1511.05030}}].
	%%CITATION = ARXIV:1511.05030;%%.
	
	\bibitem{Hambye:2008bq}
	T.~Hambye, {\it {Hidden vector dark matter}},
	\href{http://dx.doi.org/10.1088/1126-6708/2009/01/028}{{\em JHEP} {\bfseries
			01} (2009) 028} [\href{http://arxiv.org/abs/0811.0172}{{\ttfamily
			arXiv:0811.0172}}].
	%%CITATION = ARXIV:0811.0172;%%.
	
	\bibitem{Baek:2012se}
	S.~Baek, P.~Ko, W.-I. Park, and E.~Senaha, {\it {Higgs Portal Vector Dark
			Matter : Revisited}},
	\href{http://dx.doi.org/10.1007/JHEP05(2013)036}{{\em JHEP} {\bfseries 05}
		(2013) 036} [\href{http://arxiv.org/abs/1212.2131}{{\ttfamily
			arXiv:1212.2131}}].
	%%CITATION = ARXIV:1212.2131;%%.
	
	\bibitem{Farzan:2012hh}
	Y.~Farzan and A.~R. Akbarieh, {\it {VDM: A model for Vector Dark Matter}},
	\href{http://dx.doi.org/10.1088/1475-7516/2012/10/026}{{\em JCAP} {\bfseries
			1210} (2012) 026} [\href{http://arxiv.org/abs/1207.4272}{{\ttfamily
			arXiv:1207.4272}}].
	%%CITATION = ARXIV:1207.4272;%%.
	
	\bibitem{Carone:2013wla}
	C.~D. Carone and R.~Ramos, {\it {Classical scale-invariance, the electroweak
			scale and vector dark matter}},
	\href{http://dx.doi.org/10.1103/PhysRevD.88.055020}{{\em Phys. Rev.} {\bfseries
			D88} (2013) 055020} [\href{http://arxiv.org/abs/1307.8428}{{\ttfamily
			arXiv:1307.8428}}].
	%%CITATION = ARXIV:1307.8428;%%.
	
	\bibitem{Hambye:2013sna}
	T.~Hambye and A.~Strumia, {\it {Dynamical generation of the weak and Dark
			Matter scale}},
	\href{http://dx.doi.org/10.1103/PhysRevD.88.055022}{{\em Phys. Rev.} {\bfseries
			D88} (2013) 055022} [\href{http://arxiv.org/abs/1306.2329}{{\ttfamily
			arXiv:1306.2329}}].
	%%CITATION = ARXIV:1306.2329;%%.
	
	\bibitem{Chiang:2013kqa}
	C.-W. Chiang, T.~Nomura, and J.~Tandean, {\it {Non-abelian Dark Matter with
			Resonant Annihilation}},
	\href{http://dx.doi.org/10.1007/JHEP01(2014)183}{{\em JHEP} {\bfseries 01}
		(2014) 183} [\href{http://arxiv.org/abs/1306.0882}{{\ttfamily
			arXiv:1306.0882}}].
	%%CITATION = ARXIV:1306.0882;%%.
	
	\bibitem{Baek:2014goa}
	S.~Baek, P.~Ko, W.-I. Park, and Y.~Tang, {\it {Indirect and direct signatures
			of Higgs portal decaying vector dark matter for positron excess in cosmic
			rays}},
	\href{http://dx.doi.org/10.1088/1475-7516/2014/06/046}{{\em JCAP} {\bfseries
			1406} (2014) 046} [\href{http://arxiv.org/abs/1402.2115}{{\ttfamily
			arXiv:1402.2115}}].
	%%CITATION = ARXIV:1402.2115;%%.
	
	\bibitem{Chen:2015dea}
	C.-H. Chen and T.~Nomura, {\it {Searching for vector dark matter via Higgs
			portal at the LHC}},
	\href{http://dx.doi.org/10.1103/PhysRevD.93.074019}{{\em Phys. Rev.} {\bfseries
			D93} no.~7, (2016) 074019} [\href{http://arxiv.org/abs/1507.00886}{{\ttfamily
			arXiv:1507.00886}}].
	%%CITATION = ARXIV:1507.00886;%%.
	
	\bibitem{DiChiara:2015bua}
	S.~Di~Chiara and K.~Tuominen, {\it {A minimal model for SU(N ) vector dark
			matter}},
	\href{http://dx.doi.org/10.1007/JHEP11(2015)188}{{\em JHEP} {\bfseries 11}
		(2015) 188} [\href{http://arxiv.org/abs/1506.03285}{{\ttfamily
			arXiv:1506.03285}}].
	%%CITATION = ARXIV:1506.03285;%%.
	
	\bibitem{Gross:2015cwa}
	C.~Gross, O.~Lebedev, and Y.~Mambrini, {\it {Non-Abelian gauge fields as dark
			matter}},
	\href{http://dx.doi.org/10.1007/JHEP08(2015)158}{{\em JHEP} {\bfseries 08}
		(2015) 158} [\href{http://arxiv.org/abs/1505.07480}{{\ttfamily
			arXiv:1505.07480}}].
	%%CITATION = ARXIV:1505.07480;%%.
	
	\bibitem{Karam:2016rsz}
	A.~Karam and K.~Tamvakis,
	{\it {Dark Matter from a Classically Scale-Invariant $SU(3)_X$}},
	[\href{http://arxiv.org/abs/1607.01001}{{\ttfamily arXiv:1607.01001}}].
	%%CITATION = ARXIV:1607.01001;%%.
	
	\bibitem{Forestell:2016qhc}
	L.~Forestell, D.~E. Morrissey, and K.~Sigurdson,
	{\it {Non-Abelian Dark Forces and the Relic Densities of Dark Glueballs}},
	[\href{http://arxiv.org/abs/1605.08048}{{\ttfamily arXiv:1605.08048}}].
	%%CITATION = ARXIV:1605.08048;%%.
	
	\bibitem{Kribs:2016cew}
	G.~D. Kribs and E.~T. Neil, {\it {Review of strongly-coupled composite dark
			matter models and lattice simulations}},
	\href{http://dx.doi.org/10.1142/S0217751X16430041}{{\em Int. J. Mod. Phys.}
		{\bfseries A31} no.~22, (2016) 1643004}
	[\href{http://arxiv.org/abs/1604.04627}{{\ttfamily arXiv:1604.04627}}].
	%%CITATION = ARXIV:1604.04627;%%.
	
	\bibitem{Soni:2016gzf}
	A.~Soni and Y.~Zhang, {\it {Hidden SU(N) Glueball Dark Matter}},
	\href{http://dx.doi.org/10.1103/PhysRevD.93.115025}{{\em Phys. Rev.} {\bfseries
			D93} no.~11, (2016) 115025}
	[\href{http://arxiv.org/abs/1602.00714}{{\ttfamily arXiv:1602.00714}}].
	%%CITATION = ARXIV:1602.00714;%%.
	
	\bibitem{Yamanaka:2014pva}
	N.~Yamanaka, S.~Fujibayashi, S.~Gongyo, and H.~Iida,
	{\it {Dark matter in the hidden gauge theory}},
	[\href{http://arxiv.org/abs/1411.2172}{{\ttfamily arXiv:1411.2172}}].
	%%CITATION = ARXIV:1411.2172;%%.
	
	\bibitem{Boddy:2014yra}
	K.~K. Boddy, J.~L. Feng, M.~Kaplinghat, and T.~M.~P. Tait, {\it
		{Self-Interacting Dark Matter from a Non-Abelian Hidden Sector}},
	\href{http://dx.doi.org/10.1103/PhysRevD.89.115017}{{\em Phys. Rev.} {\bfseries
			D89} no.~11, (2014) 115017} [\href{http://arxiv.org/abs/1402.3629}{{\ttfamily
			arXiv:1402.3629}}].
	%%CITATION = ARXIV:1402.3629;%%.
	
	\bibitem{Faraggi:2000pv}
	A.~E. Faraggi and M.~Pospelov, {\it {Self-interacting dark matter from the
			hidden heterotic string sector}},
	\href{http://dx.doi.org/10.1016/S0927-6505(01)00121-9}{{\em Astropart. Phys.}
		{\bfseries 16} (2002) 451--461}
	[\href{http://arxiv.org/abs/hep-ph/0008223}{{\ttfamily hep-ph/0008223}}].
	%%CITATION = HEP-PH/0008223;%%.
	
	\bibitem{Hambye:2009fg}
	T.~Hambye and M.~H.~G. Tytgat, {\it {Confined hidden vector dark matter}},
	\href{http://dx.doi.org/10.1016/j.physletb.2009.11.050}{{\em Phys. Lett.}
		{\bfseries B683} (2010) 39--41}
	[\href{http://arxiv.org/abs/0907.1007}{{\ttfamily arXiv:0907.1007}}].
	%%CITATION = ARXIV:0907.1007;%%.
	
	\bibitem{Boehm:2004th}
	C.~Boehm and R.~Schaeffer, {\it {Constraints on dark matter interactions from
			structure formation: Damping lengths}},
	\href{http://dx.doi.org/10.1051/0004-6361:20042238}{{\em Astron. Astrophys.}
		{\bfseries 438} (2005) 419--442}
	[\href{http://arxiv.org/abs/astro-ph/0410591}{{\ttfamily astro-ph/0410591}}].
	%%CITATION = ASTRO-PH/0410591;%%.
	
	\bibitem{Green:2005fa}
	A.~M. Green, S.~Hofmann, and D.~J. Schwarz, {\it {The First wimpy halos}},
	\href{http://dx.doi.org/10.1088/1475-7516/2005/08/003}{{\em JCAP} {\bfseries
			0508} (2005) 003} [\href{http://arxiv.org/abs/astro-ph/0503387}{{\ttfamily
			astro-ph/0503387}}].
	%%CITATION = ASTRO-PH/0503387;%%.
	
	\bibitem{Loeb:2005pm}
	A.~Loeb and M.~Zaldarriaga, {\it {The Small-scale power spectrum of cold dark
			matter}},
	\href{http://dx.doi.org/10.1103/PhysRevD.71.103520}{{\em Phys. Rev.} {\bfseries
			D71} (2005) 103520} [\href{http://arxiv.org/abs/astro-ph/0504112}{{\ttfamily
			astro-ph/0504112}}].
	%%CITATION = ASTRO-PH/0504112;%%.
	
	\bibitem{Carroll:mha}
	L.~Ackerman, M.~R. Buckley, S.~M. Carroll, and M.~Kamionkowski, {\it {Dark
			Matter and Dark Radiation}},
	\href{http://dx.doi.org/10.1103/PhysRevD.79.023519}{{\em Phys. Rev.} {\bfseries
			D79} (2009) 023519} [\href{http://arxiv.org/abs/0810.5126}{{\ttfamily
			arXiv:0810.5126}}].
	%%CITATION = ARXIV:0810.5126;%%.
	
	\bibitem{Baek:2013qwa}
	S.~Baek, P.~Ko, and W.-I. Park, {\it {Singlet Portal Extensions of the Standard
			Seesaw Models to a Dark Sector with Local Dark Symmetry}},
	\href{http://dx.doi.org/10.1007/JHEP07(2013)013}{{\em JHEP} {\bfseries 07}
		(2013) 013} [\href{http://arxiv.org/abs/1303.4280}{{\ttfamily
			arXiv:1303.4280}}].
	%%CITATION = ARXIV:1303.4280;%%.
	
	\bibitem{Gondolo:1990dk}
	P.~Gondolo and G.~Gelmini, {\it {Cosmic abundances of stable particles:
			Improved analysis}},
	\href{http://dx.doi.org/10.1016/0550-3213(91)90438-4}{{\em Nucl. Phys.}
		{\bfseries B360} (1991) 145--179}.
	%%CITATION = NUPHA,B360,145;%%.
	
	\bibitem{Giudice:2000ex}
	G.~F. Giudice, E.~W. Kolb, and A.~Riotto, {\it {Largest temperature of the
			radiation era and its cosmological implications}},
	\href{http://dx.doi.org/10.1103/PhysRevD.64.023508}{{\em Phys. Rev.} {\bfseries
			D64} (2001) 023508} [\href{http://arxiv.org/abs/hep-ph/0005123}{{\ttfamily
			hep-ph/0005123}}].
	%%CITATION = HEP-PH/0005123;%%.
	
	\bibitem{Scherrer:1984fd}
	R.~J. Scherrer and M.~S. Turner, {\it {Decaying Particles Do Not Heat Up the
			Universe}},
	\href{http://dx.doi.org/10.1103/PhysRevD.31.681}{{\em Phys. Rev.} {\bfseries
			D31} (1985) 681}.
	%%CITATION = PHRVA,D31,681;%%.
	
	\bibitem{class}
	D.~Blas, J.~Lesgourgues, and T.~Tram, {\it {The Cosmic Linear Anisotropy
			Solving System (CLASS) II: Approximation schemes}},
	\href{http://dx.doi.org/10.1088/1475-7516/2011/07/034}{{\em JCAP} {\bfseries
			1107} (2011) 034} [\href{http://arxiv.org/abs/1104.2933}{{\ttfamily
			arXiv:1104.2933}}].
	%%CITATION = ARXIV:1104.2933;%%.
	
	\bibitem{Ma:1995ey}
	C.-P. Ma and E.~Bertschinger, {\it {Cosmological perturbation theory in the
			synchronous and conformal Newtonian gauges}},
	\href{http://dx.doi.org/10.1086/176550}{{\em Astrophys. J.} {\bfseries 455}
		(1995) 7--25} [\href{http://arxiv.org/abs/astro-ph/9506072}{{\ttfamily
			astro-ph/9506072}}].
	%%CITATION = ASTRO-PH/9506072;%%.
	
\end{thebibliography}
\providecommand{\href}[2]{#2}\begingroup\raggedright\endgroup

\end{document}